\documentclass[acmsmall,screen]{acmart}

\AtBeginDocument{%
  }

\usepackage{amsmath,amssymb,amsfonts}
\usepackage{algorithmic}
\usepackage{graphicx}
\usepackage{textcomp}
\usepackage{xcolor}
\newboolean{showcomments}
\setboolean{showcomments}{true}
\usepackage{adjustbox}
\usepackage{algorithm}
\usepackage{algorithmic}
\usepackage{amsmath,amsfonts}
\usepackage{array}
\usepackage{balance}
\usepackage{booktabs}
\usepackage[skip=1pt,labelfont=bf]{caption}
\usepackage{calc}
\usepackage{calligra}
\usepackage{color}
\usepackage{colortbl}
\usepackage{courier}
\usepackage{csvsimple}
\usepackage{enumitem}
\usepackage{fancybox}
\usepackage{fontenc}
\usepackage{graphicx}
\usepackage{listings}
\usepackage{longtable}
\usepackage{lscape}
\usepackage{makecell}
\usepackage{marvosym}
\usepackage{moreverb}
\usepackage{multicol}
\usepackage{multirow}
\usepackage{pifont}
\usepackage{pgfplots}
\usepackage{pgf-pie}
\usepackage{rotating}
\usepackage{setspace}
\usepackage{siunitx}
\usepackage{soul}
\usepackage{subcaption}
\usepackage{tablefootnote}
\usepackage[most]{tcolorbox}
\usepackage{microtype}
\usepackage{threeparttable}
\usepackage{tikz}
\usepackage[normalem]{ulem}
\usepackage{url}
\usepackage{wasysym}
\usepackage{xspace}

\usepgfplotslibrary{statistics}
\usetikzlibrary{matrix, shapes.geometric, arrows}
\usetikzlibrary{shapes, arrows, positioning}

\newcolumntype{L}[1]{>{\raggedright\let\newline\\\arraybackslash\hspace{0pt}}m{#1}}
\newcolumntype{C}[1]{>{\centering\let\newline\\\arraybackslash\hspace{0pt}}m{#1}}
\newcolumntype{R}[1]{>{\raggedleft\let\newline\\\arraybackslash\hspace{0pt}}m{#1}}

\definecolor{codegreen}{rgb}{0,0.6,0}
\definecolor{codered}{rgb}{1,0,0}
\definecolor{codegray}{rgb}{0.5,0.5,0.5}
\definecolor{codepurple}{rgb}{0.58,0,0.82}
\definecolor{backcolour}{rgb}{0.95,0.95,0.92}
\definecolor{lightgray}{gray}{0.9}

\definecolor{DarkOrange}{rgb}{0.8,0.3,0.0}
\definecolor{DarkCyan}{rgb}{0.0, 0.55, 0.55}
\definecolor{DarkCyel}{rgb}{1.0, 0.49, 0.0}
\definecolor{yellow-green}{rgb}{0.6, 0.8, 0.2}

\newcolumntype{?}{!{\vrule width 1pt}}

\newcommand{\tool}{\textsc{KGCompass}\xspace}

\newcommand{\find}[1]{
\begin{tcolorbox}[leftrule=1mm,toprule=0mm,bottomrule=0mm,left=1pt,right=2pt,top=2pt,bottom=2pt]
\em #1
\end{tcolorbox}
}

\lstdefinelanguage{mymarkdown}{
    morekeywords={*,\#, \#\#, \#\#\#},
    sensitive=false,
    morecomment=[l]{//},
    morestring=[b]",
    commentstyle=\color{codegreen},
    keywordstyle=\color{magenta},
    numberstyle=\tiny\color{codegray},
    stringstyle=\color{codepurple},
    basicstyle=\tiny,
    breakatwhitespace=false,         
    breaklines=true,
    breakindent=0pt,
    keepspaces=true,                 
    numbers=left,                    
    numbersep=5pt,                  
    showspaces=false,                
    showstringspaces=false,
    showtabs=false,                  
    tabsize=2,
}

\lstdefinestyle{mystyle}{
    commentstyle=\color{codegreen},
    keywordstyle=\color{magenta},
    numberstyle=\small\color{black},
    stringstyle=\color{codepurple},
    basicstyle=\scriptsize\ttfamily,
    breakatwhitespace=false,
    breaklines=true,
    captionpos=b,
    keepspaces=true,
    showspaces=false,
    showstringspaces=false,
    showtabs=false,
    tabsize=2
}

\lstdefinelanguage{diff}{
  morecomment=[f][\color{blue}]{@@},     
  morecomment=[f][\color{red}]-,         
  morecomment=[f][\color{codegreen}]+,       
  morecomment=[f][\color{red}]{---}, 
  morecomment=[f][\color{codegreen}]{+++},
  numberstyle=\tiny\color{codegray},
  numbers=left,                    
  numbersep=5pt,         
}

\setlist{noitemsep} 

\definecolor{darkpastelred}{rgb}{0.76, 0.23, 0.13}
\definecolor{ao(english)}{rgb}{0.0, 0.5, 0.0}

\definecolor{darkpastelred}{rgb}{0.76, 0.23, 0.13}
\definecolor{ao(english)}{rgb}{0.0, 0.5, 0.0}

\newboolean{useblue}
\setboolean{useblue}{false} 

\newcommand{\maybeblue}[1]{%
    \ifthenelse{\boolean{useblue}}%
    {\textcolor{blue}{#1}}%
    {#1}%
}
\begin{document}
\author{Boyang Yang}
\author{Jiadong Ren}
\author{Shunfu Jin}
\affiliation{%
  \institution{Yanshan University}
  \country{China}
}

\author{Yang Liu}
\affiliation{%
  \institution{Nanyang Technological University}
  \country{Singapore}
}

\author{Feng Liu}
\author{Bach Le}
\affiliation{%
  \institution{University of Melbourne}
  \country{Australia}
}

\author{Haoye Tian}
\authornote{Corresponding author.}
\affiliation{%
  \institution{Aalto University}
  \country{Finland}
}
\definecolor{DarkOrange}{rgb}{0.8,0.3,0.0}
\definecolor{DarkCyan}{rgb}{0.0, 0.55, 0.55}
\definecolor{DarkCyel}{rgb}{1.0, 0.49, 0.0}
\definecolor{yellow-green}{rgb}{0.6, 0.8, 0.2}

\newcommand{\todoc}[2]{{\textcolor{#1} {\textbf{#2}}}}
\newcommand{\todoblue}[1]{\todoc{blue}{\textbf{#1}}}
\newcommand{\todogreen}[1]{\todoc{yellow-green}{\textbf{#1}}}
\newcommand{\todored}[1]{\todoc{red}{\textbf{#1}}}
\newcommand{\bachle}[1]{\mynote{Bach}{\todoblue{#1}}}
\newcommand{\tian}[1]{\mynote{Haoye}{\todored{#1}}}
\newcommand{\yang}[1]{\mynote{Boyang}{\todogreen{#1}}}

\title{Enhancing repository-level software repair via repository-aware knowledge graphs} 

\begin{abstract}
Repository-level software repair faces challenges in bridging semantic gaps between issue descriptions and code patches. Existing approaches, which primarily rely on large language models (LLMs), are hindered by semantic ambiguities, limited understanding of structural context, and insufficient reasoning capabilities. To address these limitations, we propose \tool with two innovations: (1) a novel repository-aware knowledge graph (KG) that accurately links repository artifacts (issues and pull requests) and codebase entities (files, classes, and functions), allowing us to effectively narrow down the vast search space to only 20 most relevant functions with accurate candidate fault locations and contextual information, and (2) a path-guided repair mechanism that leverages KG-mined entity paths, tracing through which allows us to augment LLMs with relevant contextual information to generate precise patches along with their explanations. Experimental results in the SWE-bench Lite demonstrate that \tool achieves state-of-the-art single-LLM repair performance (58.3\%) and function-level fault location accuracy (56.0\%) across open-source approaches with a single repair model, costing only \$0.2 per repair. Among the bugs that \tool successfully localizes, 89.7\% lack explicit location hints in the issue and are found only through multi-hop graph traversal, where pure LLMs struggle to locate bugs accurately. Relative to pure-LLM baselines, \tool lifts the resolved rate by 50.8\% on Claude-4 Sonnet, 30.2\% on Claude-3.5 Sonnet, 115.7\% on DeepSeek-V3, and 156.4\% on Qwen2.5 Max. These consistent improvements demonstrate that this graph-guided repair framework delivers model-agnostic, cost-efficient repair and sets a strong new baseline for repository-level repair.
\end{abstract}

\maketitle

\section{Introduction}

Large Language Models (LLMs) have demonstrated remarkable coding capabilities in code generation and repair tasks~\cite{liu2023program,liu2020efficiency,zhu2024grammart5,lyu2024automatic,parasaram2024factselectionproblemllmbased,yang2025survey}.
To evaluate capabilities at a realistic scale, SWE-bench benchmark compiles 2,294 issues with corresponding codebases and test suites from 12 Python repositories~\cite{jimenez2024swebench}.
Because running the full benchmark is resource-intensive, the authors also released SWE-bench Lite, a curated subset of 300 tasks that maintains the original difficulty while reducing compute costs.
As a result, SWE-bench Lite has become the most widely used benchmark for repository-level repair tasks~\cite{xia2024agentless,wang2024openhandsopenplatformai, patchkitty,orwall2024moatless}.
Recent research on these benchmarks shows that the biggest challenge for LLM-based repair is to precisely pinpoint one or a few buggy functions hidden among a repository including thousands of files and more functions~\cite{yu2025orcaloca,jimenez2024swebench}. Imprecise fault location propagates to patch generation and depresses end-to-end repair accuracy.

To address these repository-level repair challenges, researchers have explored two main categories of approaches: agentic~\cite{gautam2024supercoder2,yang2024sweagent,wang2024openhandsopenplatformai,koduai,composio} and procedural~\cite{orwall2024moatless,xia2024agentless,li2025patchpilot}. Agentic approaches equip LLMs with various tools for autonomous planning and execution through multiple specialized agents, but lack interpretable reasoning and controlled decision planning. Procedural approaches mitigate these issues through expert-designed workflows, offering more precise and interpretable processes~\cite{xia2024agentless,orwall2024moatless}. A central bottleneck of procedural approaches is the selection of candidate locations provided to the patch generation. Systems such as Agentless~\cite{xia2024agentless} pass only one file location per repair; if that prediction is wrong, the repair often fails. Supplying more candidates improves recall but inflates context, increases cost, and raises hallucination risk for long inputs, which degrades patch accuracy~\cite{zhang2025llm}.
Consequently, the existing approaches still face two major limitations:

\begin{itemize}[leftmargin=*]
\item[\ding{172}] \textbf{Imprecise Fault Location}: Precise fault location at repository scale requires cross-modal grounding that aligns natural-language bug reports to codebase entities and links related issues and pull requests to those entities. In SWE-bench Lite, only about 32.7\% of issue descriptions of instances explicitly include the name of the file or the function of the ground truth location. Yet, most issues mention at least one existing repository identifier, such as an issue or PR number or a file path, which can serve as an anchor. Existing KG-based repair systems emphasize text analysis or intra-code structure and leave this alignment underspecified, such as RepoGraph~\cite{ouyang2024repograph}. Procedural repair systems like Agentless~\cite{xia2024agentless} also commit to a single predicted location rather than fusing corroborated locations across multiple files that supply repair context. This missing grounding makes precise pinpointing difficult and lowers repair accuracy on large codebases~\cite{jimenez2024swebench}.

\item[\ding{173}] \textbf{Lack of Decision Interpretability}: Repository-level software repair requires traceable reasoning chains to validate the repair decisions~\cite{antoniades2024swe}. While existing state-of-the-art approaches have achieved promising results, they lack interpretable decision processes in fault location and patch generation~\cite{rombaut2024watson}. Current agentic approaches~\cite{wang2024openhandsopenplatformai,yang2024sweagent,composio} often make decisions through complex black-box interactions without clear reasoning paths, while procedural approaches~\cite{xia2024agentless,orwall2024moatless} rank multiple candidate patches but provide limited insight into their final choices for users. On SWE-bench Lite, while state-of-the-art repair systems achieve more than 60.0\% repair accuracy, they fail to provide transparent reasoning information, especially for fault location, limiting both trustworthiness and practical application.
\end{itemize}

\textbf{This paper.} We present \tool, a novel approach that accurately links code structure with repository metadata into a unified database, namely a knowledge graph, to achieve more accurate and interpretable software repair. Given a natural language bug report and a repository codebase, our approach constructs a knowledge graph that captures relationships between repository artifacts (issues and pull requests) and codebase entities (files, functions, classes). This unified representation allows us to effectively trace the semantic connection between bug descriptions and potential fault locations, reducing the search space from thousands to 20 most relevant candidates that accurately provide both fault locations and contextual information. The useful contextual information provided by the knowledge graph is then used to help LLMs generate more accurate patches. 

Experimental results show that \tool demonstrates state-of-the-art single-LLM repair performance on SWE-bench Lite at a cost of only \$0.2 per repair. With Claude-4 Sonnet, \tool attains 58.3\% repair accuracy, a 50.8\% relative gain over the pure-LLM baseline. \tool also lifts DeepSeek-V3 by 115.7\% and Qwen2.5 Max by 156.4\%  relative to their respective pure-LLM baselines, demonstrating the generalizability of \tool across diverse backbone LLMs. \tool further achieves superior 83.6\% file-level and 56.0\% function-level fault location accuracy and provides clear reasoning chains that users can inspect. Analysis shows that 89.7\% of ground-truth patches require multi-hop connections, underscoring the value of exploiting indirect relationships by KG. Ablation studies confirm the individual contribution of each component. 

Our main contributions are:
\begin{itemize}[leftmargin=*,nosep,topsep=0pt,parsep=0pt]
\item \textbf{Repository-Aware Knowledge Graph:} We present a technique for constructing a repository-aware knowledge graph that merges code entities (files, classes, and functions) with issue and pull-request artifacts, thus capturing both structural dependencies and semantic relations across the entire repository. 
\item \textbf{KG-based Repair Approach:} We introduce \tool, a KG-guided repair framework that supplies an LLM with rich structural and semantic context from KG, enabling it to locate bugs precisely and synthesize accurate patches at repository scale.
\item \textbf{Efficient Search Space Reduction:} \tool lowers the cost to only \$0.2 per repair by using KG-based candidate selection to shrink the function search space from thousands to just 20. Top 20 candidates cover 91.0\% of file-level and 66.7\% of function-level ground truth fault locations.
\item \textbf{Comprehensive Experimental Evaluation:} Results on SWE-bench Lite demonstrate \tool's highest repair accuracy (58.3\%) and function-level fault location accuracy (56.0\%) among all open-source approaches based on Claude-4 Sonnet~\cite{wang2024openhandsopenplatformai,patchkitty,orwall2024moatless,xia2024agentless}. Compared with pure-LLM baselines, \tool yields improvements of 50.8\% on Claude-4 Sonnet, 30.2\% on Claude-3.5 Sonnet, 115.7\% on DeepSeek-V3, and 156.4\% on Qwen2.5 Max, confirming the generalization of \tool across different backbone LLMs.
Further ablation studies confirm the effectiveness of each component in our approach.
\end{itemize}


\section{Motivating Example}
\label{motivating}

We present two motivating examples from SWE-bench Lite~\cite{jimenez2024swebench}, a benchmark for evaluating repository-level software repair. The first example illustrates KG-based fault location, and the second example illustrates KG-guided repair.

\begin{figure}[h]
    \centering
    \includegraphics[width=.98\textwidth]{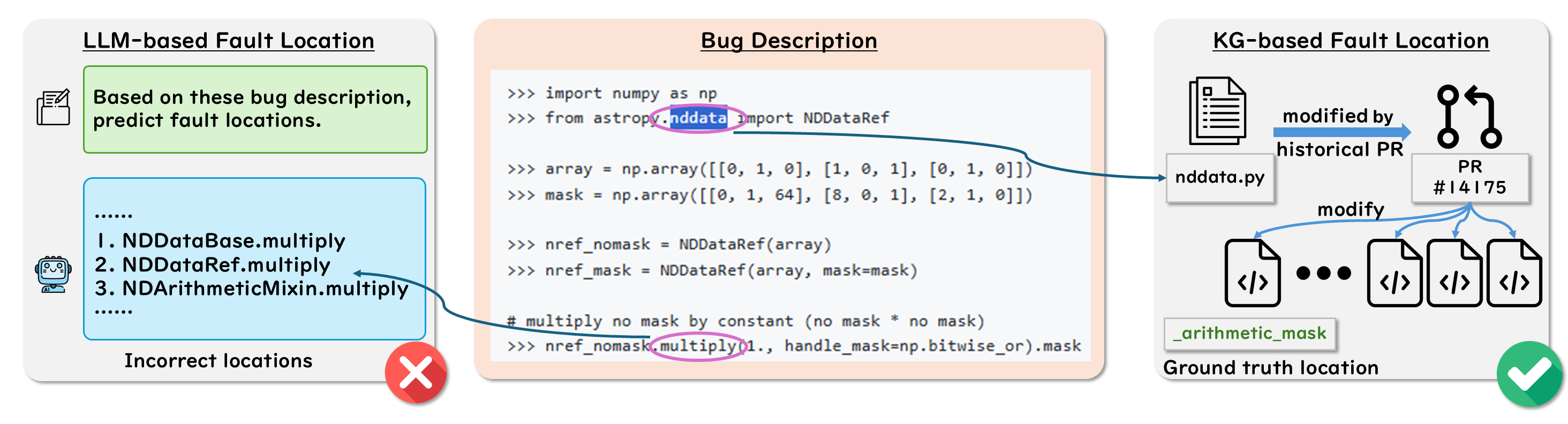} 
    \caption{Motivating Example of KG-based Fault Location}
    \label{fig:motivating_example1}
\end{figure}

\noindent\textbf{Knowledge Graph-guided Fault Location.} As shown in Figure~\ref{fig:motivating_example1}, instance astropy-14995 in SWE-bench Lite provides no explicit reference to the modified function within the ground truth patch, which only edits ``NDArithmeticMixin.\_arithmetic\_mask''. Conditioned only on the problem statement, LLM proposed three candidate locations that all revolve around a ``multiply'' routine because that token appears in the report, but none match the actual location.

Using the knowledge graph, we start from the issue text, detect the mention of ``nddata'', resolve it to the repository file ``nddata.py'', follow its link to pull request \#14175, and arrive at ``NDArithmeticMixin.\_arithmetic\_mask'', which that PR modified. Our entity relevance scoring places this function within the top five, which is sufficient for the downstream patch generation stage. This case shows how graph-derived multi-hop context counteracts lexical bias and recovers a correct location that a pure LLM misses, consistent with our KG mining and scoring strategy.

\begin{figure}[h]
    \centering
    \includegraphics[width=.98\textwidth]{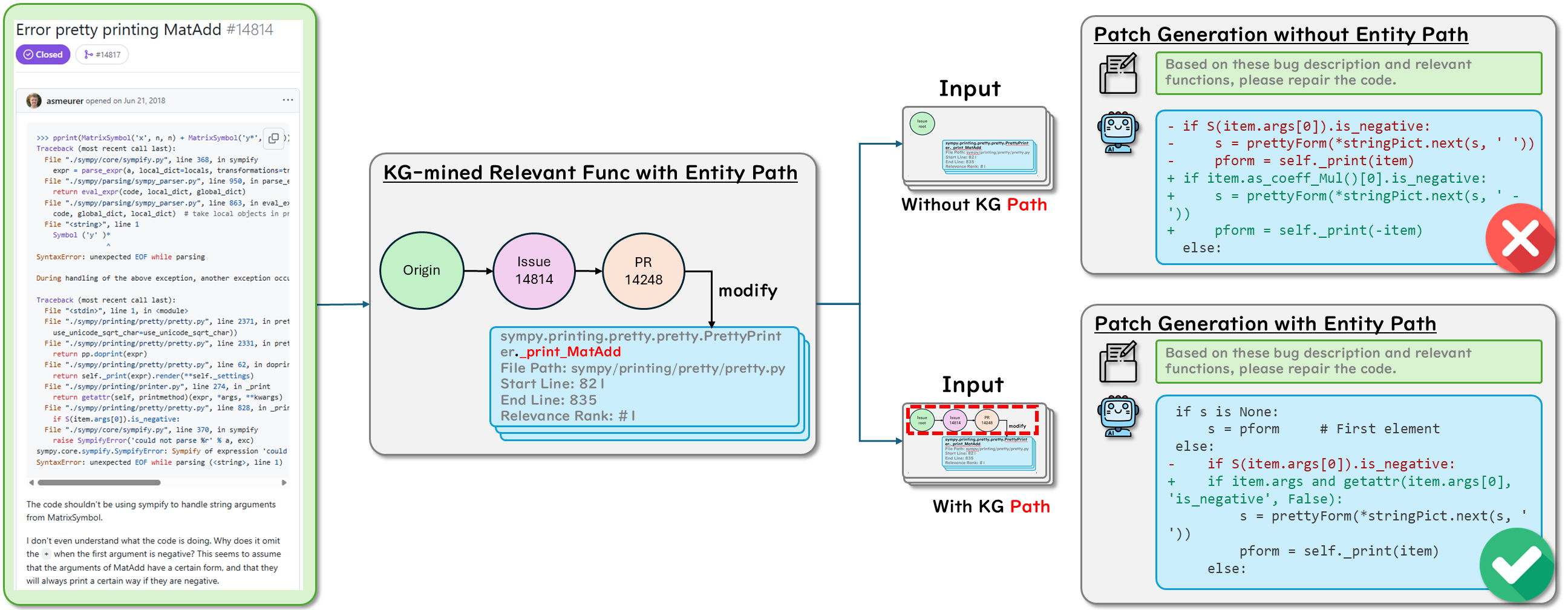} 
    \caption{Motivating Example of KG-guided Repair}
    \label{fig:motivating_example2}
\end{figure}

\noindent\textbf{Knowledge Graph-guided Repair.} As shown in Figure~\ref{fig:motivating_example2}, for the instance sympy-14814 of SWE-bench Lite, after the function ``\_print\_MatAdd'' is correctly identified, LLM produces a weak fix using ``item.as\_coeff\_Mul()[0].is\_neg ative'' that fails when processing expressions with special characters like ``y*''. However, when given the entity path (issue root → \#14814 → \#14248 → ``\_print\_MatAdd'') from the knowledge graph, the LLM generates a robust solution using ``getattr(item.args[0], 'is\_negative', False)''. The contextual information associated with this path includes comments in issue \#14814 referencing PR \#14248, suggesting that ``\_print\_MatAdd'' should use the same functions as ``\_print\_Add'' for handling plus or minus signs. This contextual information guides the LLM to adopt a more defensive programming approach rather than directly accessing attributes that might not exist. The ``getattr'' solution safely handles cases where the ``is\_negative'' attribute is unavailable, addressing the core weakness in the previous implementation. This case demonstrates how structural context from the knowledge graph helps LLMs identify relevant code patterns and relationships that would be missed when analyzing isolated code segments.

\section{Approach}
\begin{figure}[h]
    \centering
    \includegraphics[width=\textwidth]{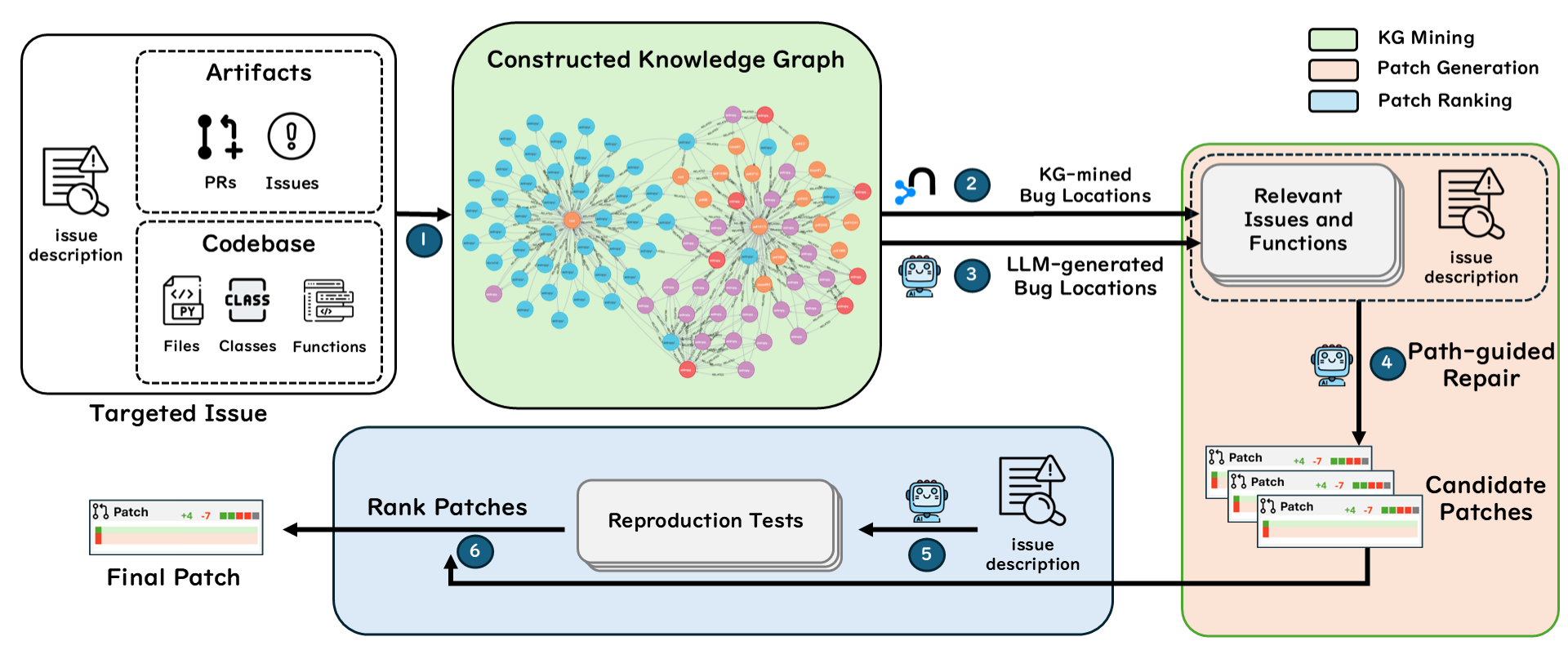}
    \caption{Overview of \tool}
    \label{fig:overview}
\end{figure}

We present \tool, a repository-level software repair framework that constructs a knowledge graph, enabling LLMs to locate accurate candidate fault locations and generate correct patches. Figure \ref{fig:overview} illustrates our approach, which unfolds in three phases: knowledge graph mining, patch generation, and patch ranking.

In the \textbf{Knowledge Graph Mining phase}, (i) we construct a comprehensive knowledge graph by integrating multiple information sources from codebases and GitHub artifacts, including issues and pull requests. (ii) Then, we identify top candidate fault locations by extracting the 15 highest-scoring functions from the graph and (iii) augmenting them with up to 5 additional candidates suggested by the LLM. In the \textbf{Patch Generation phase}, (iv) we supply the LLM with these candidate fault locations, accompanied by corresponding source code and relevant entity paths within the graph structure. In the \textbf{Patch Ranking phase}, (v) we leverage an LLM to generate reproduction tests and (vi) utilize both LLM-generated reproduction tests and regression tests within codebases to evaluate and rank the generated candidate patches, then select the final patch that is most likely to resolve the target issue.

\subsection{Knowledge Graph Mining}
\label{kg-mining}
The Knowledge Graph mining phase constructs a comprehensive knowledge graph by analyzing problem statements to extract entities and their relationships, enabling the modeling of both structural and semantic connections that are critical for accurate fault location.

We organize the knowledge graph with two groups of nodes: \emph{repository‐level artifacts} (issues and pull requests) and \emph{code‐level entities} (files, classes, and functions). For code entities, we perform a static Abstract Syntax Tree (AST) analysis that (i) records hierarchical containment edges (file$\rightarrow$class/function, class$\rightarrow$function) and (ii) derives explicit reference edges, including \emph{direct call relations} between functions, as well as import and attribute references. All such static dependencies are added to the knowledge graph as typed edges, providing the LLM with a rich structural subgraph that captures structural context such as calls, imports, attributes, and containment. For repository artifacts, we apply a template-based regular expression extractor to issue descriptions, comments, and pull request elements, identifying mentions of files, symbols, or line numbers and linking them back to the corresponding code nodes. To prevent data contamination from future artifacts, we include only artifacts whose timestamps precede the benchmark instance’s \texttt{created\_at} field, thereby mirroring the information realistically available to developers at report time. This filter rule removed 24.0\% of the initially mined related artifacts. The built knowledge graph provides both fine-grained static program dependencies and cross-domain links between natural language descriptions and code, which later serve as informative paths for fault location and patch generation.

After constructing the knowledge graph, we compute each function entity's relevance score $S(f)$ through a novel formula that combines semantic similarity with structural proximity:

\begin{equation}
\label{formula1}
S(f) = \beta^{l(f)} \cdot ( \alpha \cdot \cos_{norm}(e_i, e_f) + (1 - \alpha) \cdot \text{lev}(t_i, t_f) )
\end{equation}

Where $f$ represents a candidate function entity in the knowledge graph, $e_i$ and $e_f$ are embeddings of the problem description and function entity $f$. $t_i$ and $t_f$ are textual representations of the problem description and function entity $f$. $\text{lev}(\cdot,\cdot)$ is Levenshtein similarity normalized to $[0,1]$, $\cos_{norm}(\cdot,\cdot)$ is cosine similarity normalized to $[0, 1]$, where $\cos_{norm} = \frac{cos_{raw}+1}{2}$. $l(f)$ is the weighted shortest path length from the issue node to function entity $f$ using Dijkstra's algorithm~\cite{dijkstra}, where a smaller path length indicates a closer connection. The hyper-parameter $\alpha$ controls the balance between semantic embedding similarity and textual similarity, following the finding of~\cite{ballatore2015structural} that this combination effectively captures both contextual meaning and surface-level textual patterns. Lower $\alpha$ values prioritize surface-level textual matches, which are crucial for identifying syntactically similar but semantically distinct code elements. The path length decay factor $\beta$ determines how quickly relevance decreases with path distance through the exponential decay function $\beta^{l(f)}$. This function, adopted from prior work~\cite{zhu2016computing}, effectively models relevance decay with graph distance while preserving meaningful multi-hop relationships.

Based on these relevance scores $S(f)$, we select the top 15 functions from the knowledge graph to provide candidate fault locations and context information. We then use an LLM to identify up to 5 potential fault locations from the problem statement, complementing candidate locations with the LLM's textual understanding. This hybrid configuration combines the strengths of knowledge graph-based structural analysis and LLM-based textual understanding, creating a comprehensive set of up to 20 candidate function-level fault locations for the following patch generation phase.

\begin{figure}[h]
    \begin{lstlisting}[
        language=mymarkdown,
        style=mystyle,
        numbers=none,
        basicstyle=\scriptsize\ttfamily,
        breaklines=true,
        breakatwhitespace=false,
        columns=fullflexible
    ]
Based on the following bug description, predict potential relevant code locations:
Description:
{{problem_statement}}
Please provide a JSON array containing the predicted locations where the bug fix is needed. Each location should include the full file path and the full function name.
The function field should be one of these formats:
- `package.module.Class.function_name` # For class's functions
- `package.module.function_name` # For standalone functions
- `package.module.Class.class_level_attribute` # For class-level attributes
- `package.module.MODULE_LEVEL_VALUE` # For module-level variables
Format:
[
    {
        "file_path": "package/submodule/file.py",
        "function": "package.module.Class.function_name"
    },
    {
        "file_path": "package/other/file.py",
        "function": "package.module.MODULE_LEVEL_VALUE" 
    }
]
\end{lstlisting}
    \caption{LLM Prompt Template for Fault Location}
    \label{fig:llm-prompt}
\end{figure}

\subsection{Patch Generation}

The patch generation phase leverages the identified candidates and graph context to generate patches. \tool augments the LLM prompt with path information that illustrates structural connections (function calls, containment relationships, and references from repository artifacts). The prompt provided to the LLM consists of three key components: (1) the issue description, (2) KG-mined relevant functions with entity paths (formatted as Figure~\ref{fig:function-format}), and (3) a structured format specifying how to produce search/replace edits~\cite{xia2024agentless,gauthier2024aider}. The entity path explicitly traces connections from relevant functions to the targeted issue through intermediate entities like files, functions, issues, and pull requests. Incorporating KG-mined context enables \tool to reason about the role and relevance of each relevant function.

\begin{figure}[h]
    \begin{lstlisting}[
        language=mymarkdown,
        style=mystyle,
        numbers=none,
        basicstyle=\scriptsize\ttfamily,
        breaklines=true,
        breakatwhitespace=false,
        columns=fullflexible
    ]
### [full file path]
- signature: [namespace].[class].[function name]([params])
- relationship_path: [function reference] --([relation type])--> [entity 1] --([relation type])--> [entity 2] ... --> root
- start_line: [start]
- end_line: [end]
...
    [function code]
...
    \end{lstlisting}
    \caption{KG-mined Relevant Function Format with Entity Path for Bug Repair}
    \label{fig:function-format}
\end{figure}

To address LLMs' inherent context limitations and balance precision with diversity, \tool employs a hybrid sampling strategy, combining deterministic (temperature $=0.0$) and exploratory (temperature $>0.0$) sampling during patch generation. Deterministic sampling ensures stability and consistency, while exploratory sampling introduces diversity to cover alternative viable repair scenarios.

Generated patches commonly suffer from syntactic errors due to incorrect indentation~\cite{liu2024marscode}. To address this, \tool implements an adaptive indentation correction algorithm~(Algorithm~\ref{alg:indent})
that systematically tests minor indentation adjustments ($\pm1$, $\pm2$ levels) and selects syntactically valid variants, recovering 2\% of invalid patches. All syntactically valid patches are then passed to the subsequent patch ranking phase for evaluation through regression and reproduction tests.

\begin{algorithm}[h]
\caption{Adaptive Indentation Correction Strategy}
\label{alg:indent}
\begin{algorithmic}[1]
\STATE \textbf{INPUT:} Generated patch $P$, Original code $S$
\STATE \textbf{OUTPUT:} Syntactically valid patch $P'$ or null
\STATE $P' \leftarrow \textsc{ApplyPatch}(P, S)$
\IF{$\textsc{IsSyntaxValid}(P')$ \AND $P' \neq S$}
    \RETURN $P'$
\ENDIF

\STATE $IndentLevels \leftarrow \{-1, 1, -2, 2\}$
\FOR{$i \in IndentLevels$}
    \STATE $P_i \leftarrow \textsc{AdjustIndentation}(P, i)$
    \STATE $P' \leftarrow \textsc{ApplyPatch}(P_i, S)$
    \IF{$\textsc{IsSyntaxValid}(P')$ \AND $P' \neq S$}
        \RETURN $P'$
    \ENDIF
\ENDFOR

\RETURN null \COMMENT{No valid adjustment found}
\end{algorithmic}
\end{algorithm}

\subsection{Patch Ranking}
The patch ranking phase integrates reproduction test generation and patch prioritization to provide a comprehensive evaluation framework for selecting optimal repair solutions.

In the reproduction test generation process, \tool utilizes a prompt that combines the issue description with 20 KG-mined relevant functions. Reproduction tests are specialized test cases dynamically generated by LLMs to simulate the exact conditions and environment, verifying whether a patch addresses the proposed issue. Using DeepSeek-V3~\cite{deepseekai2024deepseekv3technicalreport}, \tool iteratively generates tests that reproduce the described issue, averaging 113 valid reproduction tests per iteration and successfully generating at least one reproduction test for 203~(67.7\%) instances in the SWE-bench Lite benchmark. All generated tests are executed within isolated Docker~\cite{docker} containers to ensure testing consistency and reproducibility.

The patch prioritization process employs a multi-layer ranking strategy. For each syntactically valid patch, we evaluate four metrics in descending priority: (1) regression test passing count, strictly limited to only those tests that the original code already passed, (2) LLM-generated reproduction test passing count, (3) majority voting (most frequently minimized LLM-generated patch), and (4) normalized patch size. The highest-ranked patch is selected as the final patch submission.

\section{Experimental Setup}
\label{experimental_setup}
\subsection{Benchmark}
We evaluate \tool on the repository-level repair benchmark SWE-bench Lite~\cite{jimenez2024swebench}. This benchmark contains 300 real GitHub issues from 12 active Python projects, together with full codebase snapshots and test suites. Each issue is presented in natural language, so the repair system must first identify the fault location within the repository and then generate a corresponding patch. SWE-bench Lite is the most popular repository-level repair benchmark currently, and it is more challenging than earlier benchmarks. Function-level repair benchmarks, such as TutorCode~\cite{cref} and HumanEval-Java~\cite{humanevaljava}, isolate the buggy code in a single file. More complex benchmarks like Defects4J~\cite{defects4j} and BugsInPy~\cite{bugsinpy} still have some limitations. Bugs in Defects4J usually appear in a single file and have a median patch size of 4 lines. At the same time, BugsInPy provides an explicit failing test for every bug and has not been updated since 2020, which increases the risk of data leakage and causes frequent environment mismatches. In SWE-bench Lite, by contrast, a tool must bridge the semantic gap between an issue text and a codebase that can span thousands of files. For example, one snapshot of the Django project within this benchmark comprises 2,750 Python files and 27,867 functions, underscoring the scale of the fault location challenge. This mix of size, realism, and lack of location hints makes SWE-bench Lite the best available benchmark for modern, multi-file, issue-driven software repair.

\subsection{Metrics}
\label{sec:metrics}
We follow standard evaluation practices for automated program repair and fault location~\cite{xia2024agentless,yu2025orcaloca,ouyang2024repograph,ma2024repository} and adopt a \emph{top‑1 success} protocol consistent with SWE-bench Lite evaluations~\cite{jimenez2024swebench}. We report four metrics: \textbf{\% Resolved}, \textbf{File Acc.}, \textbf{Func Acc.}, and \textbf{Avg Cost} per bug. \% Resolved is the fraction of instances whose top-ranked patch passes all ground-truth tests. File Acc. measures file-level fault location precision. Func Acc. measures function-level precision while resolving naming ambiguity. Avg Cost per bug captures computational efficiency under a fixed transparent workflow. Accurate fault location metrics remain a practical challenge~\cite{xia2024agentless,yu2025orcaloca}.

We compute per-instance overlaps with the Jaccard index. For files,
\begin{equation}
a_i^{\text{file}} = \frac{|F_i^\star \cap \widehat{F}_i|}{|F_i^\star \cup \widehat{F}_i|}.
\end{equation}

For code elements, we unify functions and classes into one set,
\begin{equation}
a_i^{\text{func}} = \frac{|E_i^\star \cap \widehat{E}_i|}{|E_i^\star \cup \widehat{E}_i|}.
\end{equation}

Averaged accuracies over the evaluated set $S$ are

\begin{equation}
\mathrm{FileAcc} = \frac{1}{|S|} \sum_{i \in S} a_i^{\text{file}},
\qquad
\mathrm{FuncAcc} = \frac{1}{|S|} \sum_{i \in S} a_i^{\text{func}}.
\end{equation}

$S$ includes all instances within the repository-level repair benchmark SWE-bench Lite.

Jaccard-based formulation penalizes missing and spurious targets in a single score. Prior SWE-bench Lite fault location works typically report binary match rates at the file or function level rather than IoU~\cite{yu2025orcaloca}. IoU has been used to evaluate multi-location fault location in other settings, for example, concurrency fault location~\cite{feng2025deep}. While successful fixes can occur at locations different from the ground truth patch, overlap with ground truth patches yields stable metrics that correlate with repair accuracy~\cite{xia2024agentless,yu2025orcaloca}.

Importantly, ``Func Acc.'' can be lower than ``Resolved''. The test-based success criterion tolerates extraneous edits that do not affect correctness. On SWE-bench Lite, the ground-truth patch edits one file, but a model may still modify additional functions within that file or repair the bug by changing locations that differ from the ground truth patch. Such displaced edits reduce the Jaccard overlap and therefore depress ``File Acc.'' and ``Func Acc.''. In these cases, ``File Acc.'' and ``Func Acc.'' quantify fault location alignment with the ground truth patch, while ``Resolved'' captures end-to-end repair correctness.

\subsection{Implementation Details}
\label{implementation}
We use neo4j~\cite{neo4jmain} for building the knowledge graph with plugins apoc-4.4~\cite{neo4j_apoc} and gds-2.6.8~\cite{neo4j_gds}. \tool integrates state-of-the-art LLMs and embedding models in a task-specific manner. We vary the repair model per experiment, defaulting to Claude-4 Sonnet~\cite{anthropic2025claude} for fault location and patch generation, and to DeepSeek-V3~\cite{deepseekai2024deepseekv3technicalreport} for cost-efficient test generation. Our knowledge graph construction uses jina-embeddings-v2-base-code~\cite{günther2024jinaembeddings28192token} for semantic embeddings, which was selected because it is a widely used open-source embedding model specifically designed for mixed natural language and code content in software repositories, which can accurately capture semantic relationships between issue descriptions and code entities. We empirically determined the hyper-parameters through initial parameter exploration, setting $\beta=0.6$ for path length decay and $\alpha=0.3$ for the embedding-textual similarity balance in \tool. We employ multiple temperature settings: $1.0$ for test generation to promote diversity, $0$ for deterministic LLM-based fault location, and temperatures $0$ and $1.0$ for mixture patch generation, following previous empirical studies~\cite{qiu2024efficient,zhu2024hot}.

\subsection{Baselines}
\label{subsec:baselines}
We select the top ranked state-of-the-art repair systems on SWE-bench Lite leaderboard as baselines, including ExpeRepair~\cite{mu2025experepair}, Refact.ai Agent~\cite{refact.ai}, SWE-agent~\cite{yang2024sweagent}, DARS Agent~\cite{aggarwal2025darsdynamicactionresampling}, Lingxi~\cite{lingxi}, and OpenHands~\cite{wang2024openhandsopenplatformai}.

To isolate the effect of the knowledge graph in \tool, we add an \emph{LLM-only Fault Location} variant, named \textbf{pure-LLM}. It replaces the hybrid candidate selector in \tool, which uses the knowledge graph’s top 15 functions plus up to 5 LLM suggestions, while the pure-LLM baseline proposes up to 20 function-level candidates from the issue description using the same fault location prompt (Figure \ref{fig:llm-prompt}). After candidate selection, all subsequent steps are identical to \tool, isolating the effect of KG-based fault location and ensuring a fair comparison. This keeps the pipeline identical after the candidate step and yields a fair test of how KG-based fault location affects end-to-end repair.

\subsection{Research Questions}
\begin{itemize}[noitemsep,topsep=0pt,leftmargin=*]
\item \textbf{RQ-1}: \textbf{\textit{How effective is \tool in repository-level software repair compared to state-of-the-art approaches?}} We measure repair success, fault location accuracy, and computational cost against (i) the pure-LLM baselines of Claude-4 Sonnet, Claude-3.5 Sonnet, DeepSeek-V3, and Qwen2.5 Max, and (ii) state-of-the-art open-source repair systems. Case studies further dissect \tool’s strengths and remaining limitations.

\item \textbf{RQ-2}: \textbf{\textit{How does the knowledge graph construction within \tool contribute to fault location?}} We analyze the effectiveness of our repository-aware knowledge graph in identifying relevant functions, examining multi-hop relationships, and path structures connecting issue descriptions to fault locations.

\item \textbf{RQ-3}: \textbf{\textit{What are the impact of different components in \tool on repair performance?}} Through ablation studies, we evaluate the contributions of key components, including candidate function selection strategies, entity path information in patch generation, various patch ranking approaches, and different candidate patch pool size configurations to understand their individual effects on repair success.
\end{itemize}

\section{Experiments and Results}

\subsection{RQ-1: Effectiveness of \tool}

\noindent\textbf{[Objective]}: We aim to evaluate \tool's effectiveness in repository-level software repair by comparing its repair success, location accuracy, and cost against existing state-of-the-art approaches.

\noindent\textbf{[Experimental Design]}: The comparison baselines are detailed in Subsection \ref{subsec:baselines}. All experiments utilize the most popular repository-level repair benchmark, SWE-bench Lite~\cite{jimenez2024swebench}, which has become the standard for repository-level repair due to its widespread adoption by recent state-of-the-art systems, ensuring fair and direct comparisons. For \tool, we configure 4 backbone LLMs for patch generation, including Claude-4 Sonnet, Claude-3.5 Sonnet, DeepSeek-V3, and Qwen2.5 Max, according to Section \ref{implementation}. All systems are evaluated with four metrics: repair success (``Resolved''), file-level location accuracy (\textit{\% File Acc.}), function-level location accuracy (\textit{\% Func Acc.}), and average cost per bug (``Cost''), as detailed in Section \ref{sec:metrics}. Furthermore, we perform cross-analysis on all these Claude-3.5 Sonnet and Claude-4 Sonnet based systems separately to determine whether \tool can uniquely resolve errors that other systems cannot fix. Finally, we conduct an in-depth case study of a \tool failed repair leveraging Claude-3.5 Sonnet.

\noindent\textbf{[Experimental Results]}: Across all tools, \tool ranks third overall by \% Resolved, is the top system with one single repair model, and is also the top single-LLM repair system that uses only Claude-3.5 Sonnet. Moreover, \tool with Claude-4 Sonnet achieves the highest \textit{Func Acc.} among all functions in Table~\ref{tab:main_result}. Based on Claude-3.5 Sonnet, \tool resolves 46.0\% of the 300 bugs, outperforming OpenHands~(41.7\%) and the pure-LLM baseline (35.3\%) while keeping the average cost at \$0.2 per bug. The location precision of \tool reaches 76.7\% at the file level and 49.4\% at the function level. With Claude-4 Sonnet as the repair model, \tool attains 58.3\% resolved with 83.6\% file-level accuracy and 56.0\% function-level accuracy, exceeding other single-LLM repair systems and yielding the highest function-level precision overall.

\begin{table*}[h]
\centering
\small
\resizebox{\linewidth}{!}{ 
\begin{tabular}{lccccc}
\toprule
\textbf{System} & \textbf{Repair Model} & \textbf{Resolved} & \textbf{Cost}  & \textbf{File Acc.} & \textbf{Func Acc.}\\
\midrule
ExpeRepair & Claude-4 Sonnet \& o4-mini & 181 (60.3\%) & \textasciitilde\$2.1 & 84.7 & 52.6\\
Refact.ai Agent & Claude-3.7 Sonnet \& o4-mini & 180 (60.0\%) & N/A & 80.6 & 47.0 \\
\tool & Claude-4 Sonnet & 175 (58.3\%) & \$0.2 & 83.6 & 56.0 \\
SWE-agent & Claude-4 Sonnet & 170 (56.7\%) & \textasciitilde\$1.6 & 80.9 & 53.9 \\
ExpeRepair & Claude-3.5 Sonnet \& o3-mini & 145 (48.3\%) & \$2.1 & 80.7 & 49.6 \\
SWE-agent & Claude-3.7 Sonnet & 144 (48.0\%) & \textasciitilde\$1.6 & 79.3 & 52.2\\
DARS Agent & Claude-3.5 Sonnet \& DeepSeek-R1 & 141 (47.0\%) & \$12.24 & 78.4 & 49.3\\
\tool &  Claude-3.5 Sonnet & 137~(46.0\%) & \$0.2 & 76.7 & 49.4 \\
Lingxi & Claude-3.5 Sonnet & 128 (42.7\%) & N/A & 66.1 & 39.7 \\
OpenHands & Claude-3.5 Sonnet & 125~(41.7\%) & \$1.3 & 69.3 & 43.8 \\
\midrule
pure-LLM & Claude-4 Sonnet & 116 (38.7\%) & \$0.2 & 69.0 & 44.0 \\
pure-LLM & Claude-3.5 Sonnet & 106~(35.3\%) & \$0.2 & 72.0 & 45.3\\
\bottomrule
\end{tabular}
}
\caption{SWE-bench Lite Results of Top-tier Repair Systems Without Fine-tuning}
\label{tab:main_result}
\end{table*}

\begin{figure}[h]
    \centering
    \includegraphics[width=0.45\textwidth]{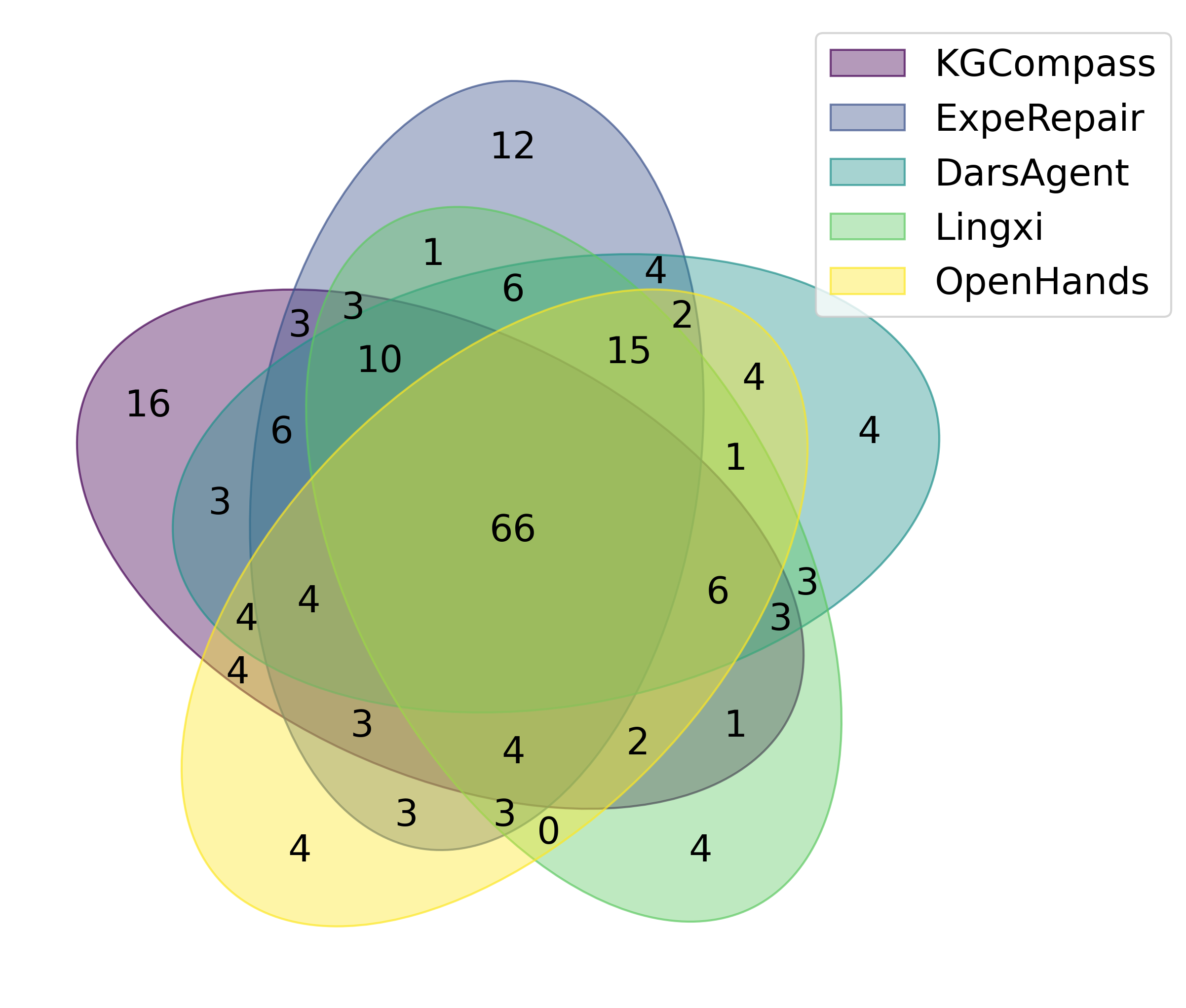}
    \caption{Intersection Analysis of \tool Against Leading Claude-3.5 based Systems}
    \label{fig:venn35}
\end{figure}

\begin{figure}[h]
    \centering
    \includegraphics[width=0.45\textwidth]{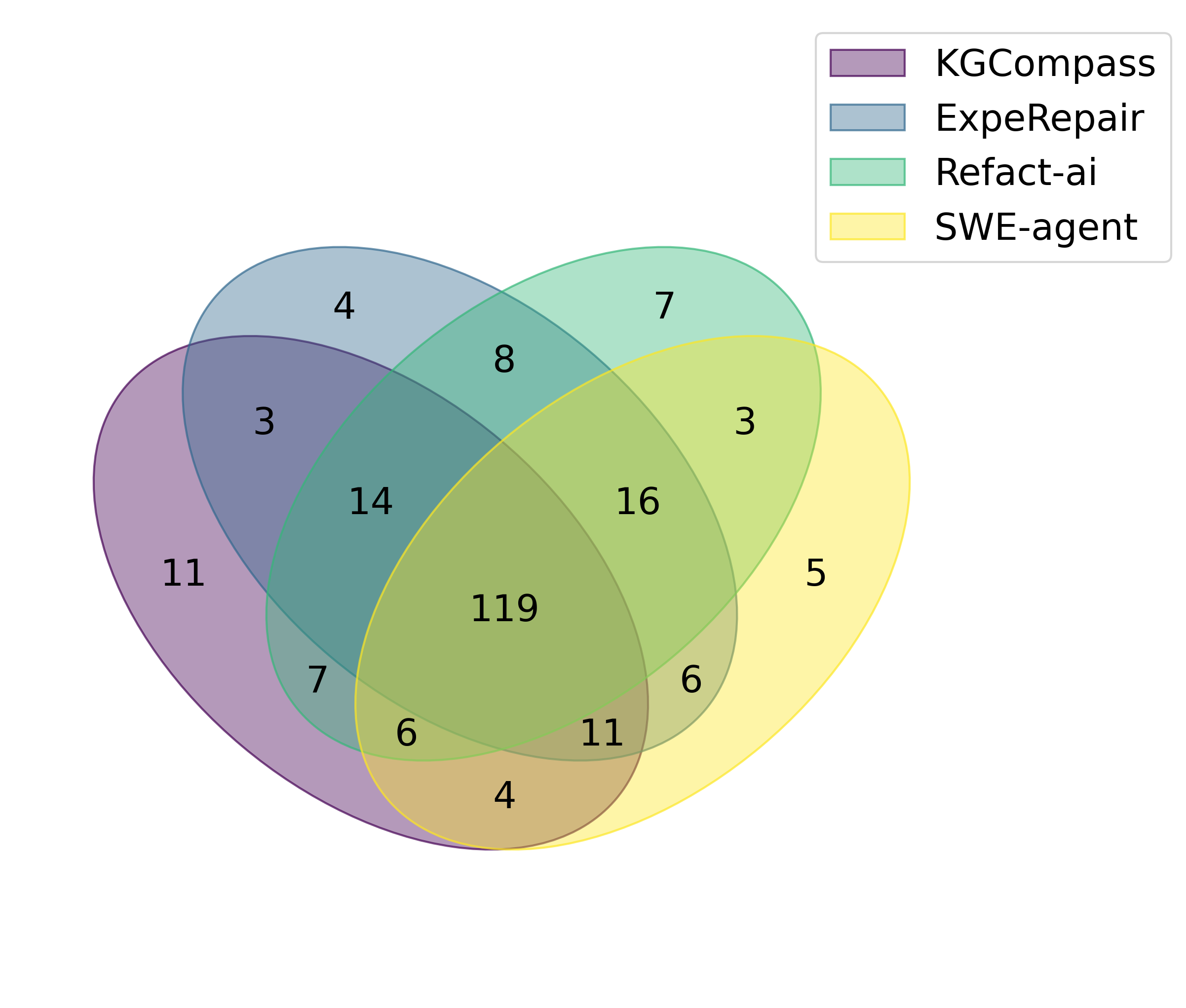}
    \caption{Intersection Analysis of \tool Against Leading Claude-4 based Systems}
    \label{fig:venn4}
\end{figure}

Figures~\ref{fig:venn35} and~\ref{fig:venn4} reveal both overlap and complementarity among the compared systems. Under the Claude-3.5 Sonnet setting, all 5 tools share 66 solved bugs. Beyond this overlap, \tool contributes the largest number of unique fixes with 16, followed by ExpeRepair with 12, and DARS Agent, Lingxi, and OpenHands with 4 each. Under the Claude-4 Sonnet setting, the shared core is 119 solved bugs. Again, \tool leads in exclusive coverage with 11 unique fixes, ahead of Refact.ai Agent with 7, SWE-agent with 5, and ExpeRepair with 4. These patterns show that the knowledge graph enables \tool to expand coverage and handle cases that other state-of-the-art systems miss.

Table \ref{tab:models_result} shows that \tool improves on every repair model. Compared with corresponding pure-LLM baselines, the resolved rate rises by 50.8\% on Claude-4 Sonnet, 30.2\% on Claude-3.5 Sonnet, 115.7\% on DeepSeek-V3, and 156.4\% on Qwen2.5 Max. File- and function-level location accuracy increases by up to 73.7\% and 139.2\%, respectively. These consistent gains indicate that the KG-enhanced repair workflow of \tool demonstrates generalizability across varying backbone LLMs.

\begin{table}[h]
\centering
\begin{tabular}{lcccc}
\toprule
\textbf{System} & \textbf{Base Model} & \textbf{Resolved} &  \textbf{ File Acc.} & \textbf{Func Acc.}\\
\midrule
\tool & Claude-4 Sonnet & 58.3\% & 83.6 & 56.0 \\
pure-LLM & Claude-4 Sonnet & 38.7\% & 69.0 & 44.0 \\
\midrule
\tool & Claude-3.5 Sonnet & 46.0\% & 76.7 & 49.4 \\
pure-LLM & Claude-3.5 Sonnet & 35.3\% & 72.0 & 45.3 \\
\midrule
\tool & DeepSeek-V3 & 36.7\%  & 76.3 & 43.7 \\
pure-LLM & DeepSeek-V3 & 17.0\% & 54.0 & 27.0 \\
\midrule
\tool & Qwen2.5 Max & 33.3\% & 74.7 & 48.6\\
pure-LLM & Qwen2.5 Max & 13.0\% & 43.0 & 20.3\\
\bottomrule
\end{tabular}
\caption{SWE-bench Lite Results of \tool vs. Pure-LLM of 4 Repair LLMs.}
\label{tab:models_result}
\end{table}

Table \ref{tab:token-cost} reports an average cost of \$0.1999 per bug. The patch phase dominates the budget at \$0.1730, split between the input prompt (\$0.1226) and the model’s response (\$0.0504). The fault location process costs \$0.0080, primarily due to the use of candidates generated by the KG, enabling the LLM to provide additional supplementary candidate locations. The repair process for each instance only needs to generate them once. Reproduction tests add \$0.0189 thanks to the lower price of DeepSeek-V3.

\begin{table}[h]
    \centering
        \begin{tabular}{lccc}
            \toprule
            \textbf{Stage} & \textbf{Tokens} & \textbf{Price \$ / M Tokens} & \textbf{Cost} \\
            \midrule
            Patch - Input & 40880.0 & \$3.00 & \$0.1226\\
            Patch - Output & 3360.9 & \$15.00 & \$0.0504\\
            Location - Input & 470.8 & \$3.00 & \$0.0014 \\
            Location - Output & 441.4 & \$15.00 & \$0.0066 \\
            Tests - Input & 63455.7 & \$0.28 & \$0.0175 \\
            Tests - Output & 2555.9 & \$0.55 & \$0.0014\\
            \midrule
            Total per Bug & & & \$0.1999 \\
            \bottomrule
        \end{tabular}
    \caption{Token and Cost Statistics of \tool}
    \label{tab:token-cost}
\end{table}

As a stratified analysis across Claude‑3.5 Sonnet and Claude‑4 Sonnet, splitting instances by whether the issue text contains an explicit file or function name shows larger gains when no hints are present and a stronger lift with Claude‑4 (Table~\ref{tab:hint_split}). With Claude‑3.5, KGCompass exceeds the pure‑LLM baseline by 18.5\% with hints (65.3\% vs 55.1\%) and by 42.4\% without hints (36.6\% vs 25.7\%). With Claude‑4, the advantage widens to 33.7\% with hints (76.5\% vs 57.2\%) and 69.5\% without hints (49.5\% vs 29.2\%). Weighted by subset shares, these splits reproduce the headline resolved rates and align with the multi‑hop patterns in Subsection~\ref{subsec:rq2}, where most successful localizations depend on indirect links captured by the knowledge graph.

\begin{table}[t]
\centering
\caption{Direct-hint vs no-hint performance on SWE-bench Lite by backbone. Baseline is the pure-LLM with the same backbone.}
\label{tab:hint_split}
\begin{tabular}{lcccc}
\toprule
\multirow{3}{*}{\textbf{Subset}} & \multicolumn{2}{c}{Claude-3.5 Sonnet} & \multicolumn{2}{c}{Claude-4 Sonnet} \\
\cmidrule(lr){2-3}\cmidrule(lr){4-5}
 & \textbf{KGCompass} & \textbf{pure-LLM} & \textbf{KGCompass} & \textbf{pure-LLM} \\
\midrule
With explicit location (32.7\%)  & \textbf{65.3\%} & 55.1\% & \textbf{76.5\%} & 57.2\% \\
Without explicit location (67.3\%) & \textbf{36.6\%} & 25.7\% & \textbf{49.5\%} & 29.2\% \\
\bottomrule
\end{tabular}
\end{table}

\noindent\textbf{[Failed Case Study]}:
Instance \texttt{django-12589} shows how the knowledge graph helps when an issue’s text provides no location hint. The bug is an ambiguous “GROUP BY” clause that triggers a runtime SQL error in Django 3.0. Pure LLM systems cannot even localize the fault, but \tool follows the path \texttt{issue $\rightarrow$ query.py $\rightarrow$ set\_group\_by} and ranks the target function first. This two-hop trace combines an issue entity, a file entity, and a function entity, a type of link that text-only analysis cannot obtain. With Claude-3.5 Sonnet, \tool produced a near-miss patch because the LLM lacked framework-specific knowledge in Django’s ORM. Using the same KG context but switching the repair model to Claude-4 Sonnet, \tool successfully repaired this instance. This supports the view that once the knowledge graph disambiguates location, stronger pretraining improves patch generation.

\find{{\bf [RQ-1] Findings:} (1) \tool ranks third overall by \% Resolved, is the best among state-of-the-art systems with a single repair model, and is also the best among systems that use only Claude-3.5 as the repair model. \tool further achieves the highest function-level location accuracy across all systems while keeping cost at \$0.2 per bug. (2) \tool yields large relative gains over pure-LLM baselines across backbones, +50.8\% on Claude-4 Sonnet, +30.2\% on Claude-3.5 Sonnet, +115.7\% on DeepSeek-V3, and +156.4\% on Qwen2.5 Max. (3) \tool uniquely fixes the most (11) bugs under Claude-4 and most (16) bugs under Claude-3.5 Sonnet, providing complementary coverage beyond existing repair systems. {\bf Insights:} (1) Leveraging a repository-aware knowledge graph to link repository artifacts and codebase entities narrows the search space and boosts repair accuracy at low cost. (2) Upgrading the repair model’s pretraining capacity further improves patch generation once the knowledge graph resolves the fault location.}

\subsection{RQ-2: Contribution of Knowledge Graph to fault location}
\label{subsec:rq2}
\noindent\textbf{[Objective]}: We aim to assess the impact of the knowledge graph on fault location by measuring its ability to (i) cover ground-truth locations and (ii) rank candidate functions accurately, thereby clarifying the knowledge graph’s role in \tool’s overall effectiveness.

\noindent\textbf{[Experimental Design]}: To assess the knowledge graph in isolation, we analyze the pure KG-mined 20 candidate bug functions, which means the list returned by the knowledge graph without any LLM-generated fault locations. First, we measure the number of ground-truth locations that fall within this set to establish coverage. Second, we examine the hop count from the starting issue node to its ground-truth function to see how often multi-hop links are needed. Third, we break down the intermediate node types on these paths to illustrate how code entities and repository artifacts are connected. Finally, we record the rank of every ground-truth function within the KG-mined 20 list to evaluate the accuracy of our relevance function (Equation~\ref{formula1}).

\begin{figure}[h]
    \centering
    \includegraphics[width=.5\columnwidth]{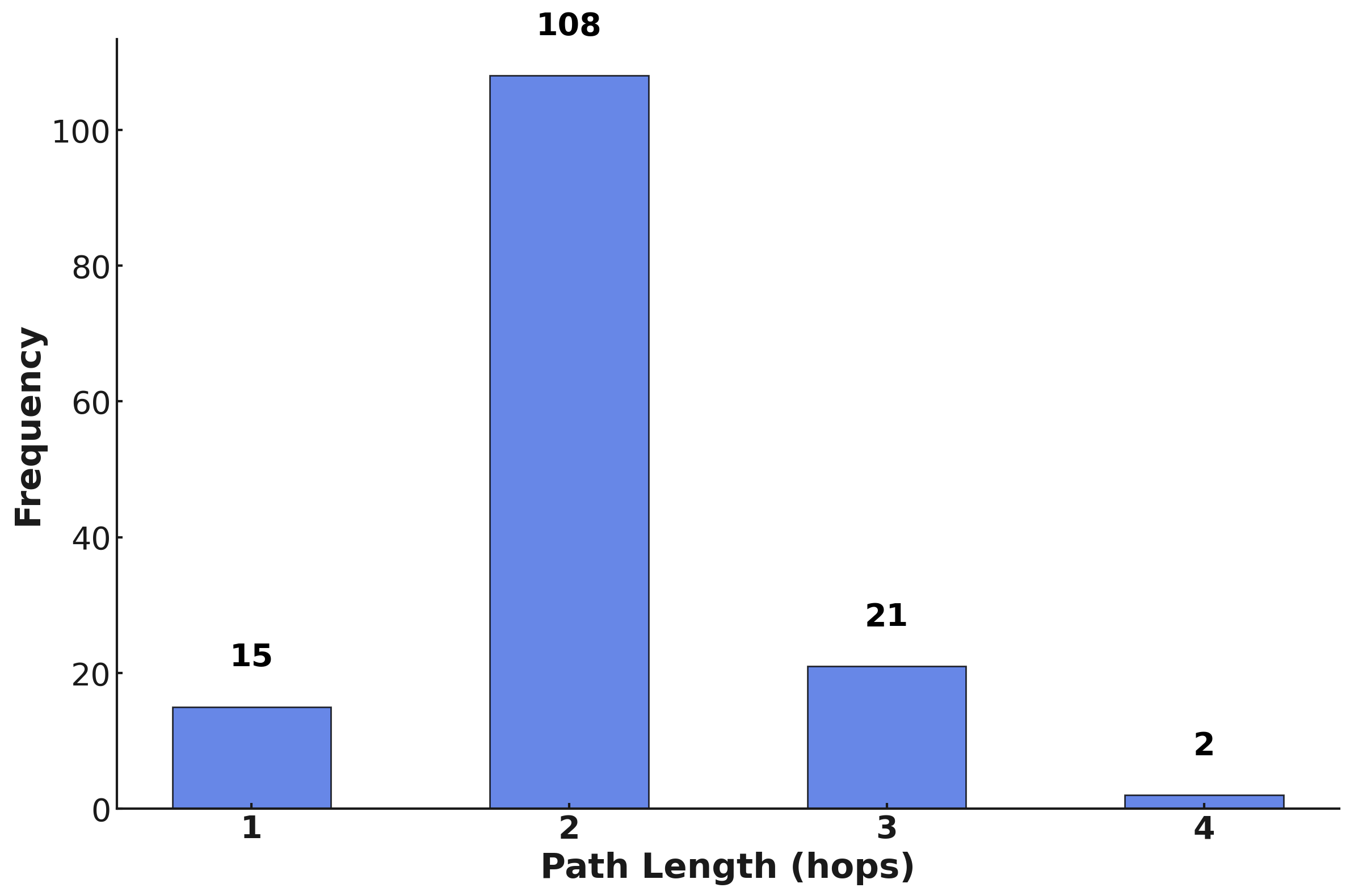}
    \caption{Path Length Distribution for Ground Truth fault locations}
    \label{fig:pathlength}
\end{figure}

\noindent\textbf{[Experimental Results]}: As shown in Table~\ref{tab:top20-loc}, KG-mined 20 relevant functions cover 68.7\% of file-level and 40.4\% of function-level ground truth fault locations, without leveraging LLMs. Figure \ref{fig:pathlength} illustrates the hop distribution of the ground truth patch from the starting issue within the knowledge graph, reveals that only 15 of the 146 ground-truth bug functions (10.3\%) are directly connected to origin issue descriptions, whereas 108 (74.0\%), 21 (14.4\%), and 2 (1.4\%) functions lie two, three, and four hops away, yielding an aggregate 89.7\% that depend on at least one intermediate entity. Some ground-truth patches modified more than one function. These statistics demonstrate that effective repository-level fault location benefits mostly from exploiting the multi-hop relations captured in the knowledge graph, since pure text-based LLM workflows display substantial model-dependent variability in accuracy, thereby validating \tool's knowledge graph-based design.

\begin{figure}[h]
    \centering
        \centering
        \begin{tikzpicture}
            \pie[
                text=legend, 
                explode={0.05, 0.3, 0.3, 0.05}, 
                radius=2.5
            ]{
                72.4/Files,
                4.5/Issues,
                7.1/Pull Requests,
                16.0/Functions
            }
        \end{tikzpicture}
    \caption{Intermediate Entity Types in KG Paths to Ground Truth Locations}
    \label{fig:kg-types}
\end{figure}

Figure \ref{fig:kg-types} further details the 156 intermediate entities observed on ground truth paths: 113 files (72.4\%), 25 functions (16.0\%), 11 pull requests (7.1\%), and 7 issues (4.5\%). The dominance of files and functions underscores the importance of static structural links mined by KG within the codebase. Meanwhile, the presence of issues and pull requests (11.6\%) demonstrates that repository artifacts provide essential context for bridging origin issue reports to the codebase. This distribution justifies the design of \tool to combine static analysis with repository artifacts in the knowledge graph for reliable location.

\begin{figure}[h]
    \centering
    \includegraphics[width=.5\columnwidth]{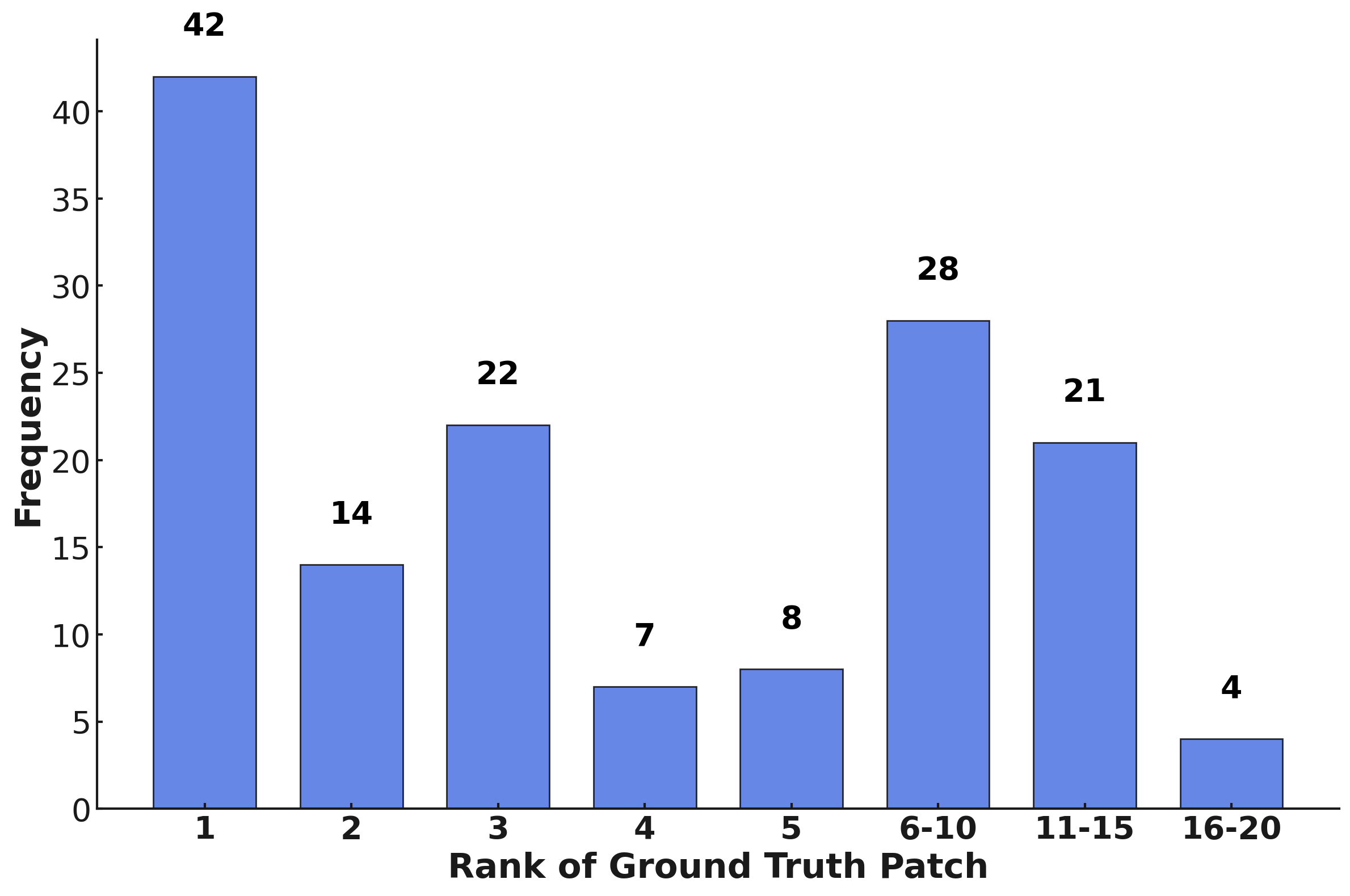}
    \caption{Rank of Ground Truth Patch in KG Candidates}
    \label{fig:gt-rank}
\end{figure}

Figure \ref{fig:gt-rank} plots the rank of each ground-truth function among the 20 KG-mined candidates. 42 functions (28.8\%) are placed at rank 1, 93 (63.7\%) fall within the first five positions, and 121 (82.9\%) appear in the top 10. There are only 4 instances beyond rank 15. The strong concentration near the top confirms the relevance score's effectiveness. In contrast, the small tail indicates that additional ranking signals are still required, thereby justifying our hybrid KG + LLM location strategy. 15 KG-mined functions with the highest relevant score hit cover 97.3\% of the obtained ground-truth locations. Therefore, we retain 15 from KG plus 5 more from LLM to provide mixed candidate function locations.

\find{{\bf [RQ-2] Findings:} (1) Only 10.3\% of bugs sit in a 1-hop relationship from the origin issue, whereas 89.7\% require 2 or more hops; accurate location, therefore, relies on traversing indirect links encoded in the knowledge graph. (2) Although files and functions form 88.4\% of intermediate nodes, the remaining 11.6\% are issues and pull requests, indicating that repository artifacts supply critical context unavailable from static code alone. (3) Within 20 KG-mined candidates, 82.9\% of ground-truth functions appear in the first 10, showing that the relevance score produces a compact, high-quality set that the hybrid KG + LLM pipeline can further refine. {\bf Insights:} By integrating the structural codebase with repository artifacts, the knowledge graph bridges the semantic gap between issue reports and buggy code, enabling reliable fault location.}

\subsection{RQ-3: Impact of Components on \tool's Repair Performance}
\noindent\textbf{[Objective]}: We measure the contribution of three key components in \tool, namely KG-based fault location, entity-path prompting, and the pool size of candidate patches.

\noindent\textbf{[Experimental Design]}: We run ablations that vary one factor at a time. All variants share the same configuration for the main pipeline: Claude-3.5 Sonnet for location and patch generation, DeepSeek-V3 for reproduction tests, identical prompt templates, and fixed temperature settings. Any change in outcome is thus attributable to the component under examination. We compare three location schemes (KG-only, LLM-only, and the hybrid KG+LLM), each restricted to 20 functions. We then test patch generation with and without KG-derived entity-path context, while fixing the candidate list and using temperature $0.0$. Next, we evaluate a cumulative ranking pipeline that starts with majority voting and successively adds regression tests, reproduction tests, and a patch-size tie-breaker, keeping previously introduced criteria. Finally, with the best ranking policy fixed, we vary the patch pool size (1, 2, 4, 6) to study the diversity–quality trade-off. In addition to the Claude-3.5 pipeline, we report Claude-4 results under the same settings to validate effectiveness and generalizability.

\noindent\textbf{[Experimental Results]}: \emph{Fault Location.} With Claude-4, the hybrid KG+LLM locator achieves 91.0\% file-level and 66.7\% function-level coverage, compared to LLM-only at 77.3\% and 52.6\%, respectively (Table~\ref{tab:top20-loc}). Claude-3.5 shows the same pattern: hybrid reaches 84.3\% file-level and 57.7\% function-level coverage, exceeding LLM-only at 81.0\% and 53.1\%, and KG-only at 68.7\% and 40.4\%. These results confirm that KG structural links and LLM semantic reasoning are complementary.

\begin{table}[h]
\centering
\begin{tabular}{lcc}
\toprule
\textbf{Strategy} & \textbf{\% GT File} & \textbf{\% GT Function} \\
\midrule
KG & 68.7 & 40.4 \\
\midrule
Claude-3.5 Sonnet & 81.0 & 53.1 \\
KG + Claude-3.5 Sonnet & 84.3 & 57.7 \\
\midrule
Claude-4 Sonnet & 77.3 & 52.6 \\
KG + Claude-4 Sonnet & 91.0 & 66.7 \\
\bottomrule
\end{tabular}
\caption{Performance Between LLM-based and Hybrid Candidate Selection Strategies}
\label{tab:top20-loc}
\end{table}

\noindent\emph{Entity-path prompting.} At temperature $0.0$ with a fixed candidate list, Claude-4 improves from 91 resolved bugs (30.3\%) without the KG to 131 (43.7\%) with KG but without entity paths, and to 142 (47.3\%) when entity paths are included (Table~\ref{tab:path_ablation}). Claude-3.5 shows the same trend, rising from 85 (28.3\%) without KG to 102 (34.0\%) without paths and to 108 (36.0\%) with paths. The motivating example illustrates why the path signal matters: only when the prompt contained the \textit{issue $\rightarrow$ PR $\rightarrow$ \texttt{\_print\_MatAdd}} path did the model produce the robust \texttt{getattr}-based patch.

\begin{table}[h]
\centering
\begin{tabular}{llc}
\toprule
\textbf{Model} & \textbf{Variant} & \textbf{Resolved}\\
\midrule
\multirow{3}{*}{Claude-3.5 Sonnet} 
 & \textbf{\tool} & \textbf{108~(36.0\%)} \\
 & w/o KG & 85~(28.3\%) \\
 & w/o Entity Path & 102~(34.0\%) \\
\midrule
\multirow{3}{*}{Claude-4 Sonnet} 
 & \textbf{\tool} & \textbf{142~(47.3\%)} \\
 & w/o KG & 91~(30.3\%) \\
 & w/o Entity Path & 131~(43.7\%) \\
\bottomrule
\end{tabular}
\caption{Ablation Study of KG and Relationship Path Information within \tool, with Temperature $0.0$}
\label{tab:path_ablation}
\end{table}

\noindent\emph{Ranking pipeline.} Strengthening the selector yields large gains and approaches the oracle ceiling. For Claude-4, accuracy rises from 139 (46.3\%) with greedy sampling to 150 (48.3\%) with majority voting, 164 (51.7\%) after adding regression tests, and 175 (58.3\%) after adding reproduction tests, while the upper bound is 182 (60.7\%) (Table~\ref{tab:patch_ranking}). Claude-3.5 follows the same progression, from 108 (36.0\%) to 121 (40.3\%), then 129 (43.0\%), and finally 138 (46.0\%), close to its 143 (47.7\%) ceiling. On Claude-4, the ground-truth upper bound reaches 182 solved bugs (60.7\%), surpassing ExpeRepair’s 181 (60.3\%) and \textbf{ranking first} among the compared systems.

\begin{table}[h]
\centering
\begin{tabular}{lcc}
\toprule
\textbf{Ranking Strategy}  & Claude-3.5 & Claude-4 \\
\midrule
Greedy Sampling  & 108~(36.0\%)  & 139 (46.3\%) \\
Majority Voting  & 121~(40.3\%) & 150 (48.3\%) \\
+ Regression tests & 129~(43.0\%) & 164 (51.7\%)\\
\ \ + Reproduction tests & 138~(46.0\%) & 175 (58.3\%) \\
\midrule
Ground-truth tests & 143~(47.7\%) & 182 (60.7\%)\\
\bottomrule
\end{tabular}
\caption{Ablation Study of Patch Ranking Strategies, with 6 Candidate Patches}
\label{tab:patch_ranking}
\end{table}

\noindent\emph{Size of Patch Pool.} Increasing the patch pool size can improve repair accuracy. With Claude-4, patch pool size with 1, 2, 4, and 6 achieve 142 (47.3\%), 148 (49.3\%), 161 (53.7\%), and 175 (58.3\%), respectively. Claude-3.5 shows the same saturation, moving from 108 (36.0\%) to 116 (38.7\%), 126 (42.0\%), and 138 (46.0\%). A larger patch pool size would likely close that gap, but it would also increase prompt length and evaluation time, driving costs up. We therefore cap the pool at 6 candidates for \tool. As inference cost drops in the future, raising the patch pool size could be a practical way to unlock the small remaining performance margin.

\begin{table}[h]
\centering
\begin{tabular}{>{\centering\arraybackslash}m{2.8cm}cccc}
\toprule
\multirow{2}{2.8cm}{\centering\textbf{Candidate Patches}} & \multicolumn{2}{c|}{\textbf{Claude3.5}} & \multicolumn{2}{c}{\textbf{Claude4}}\\
\cmidrule(lr){2-3}\cmidrule(lr){4-5}
 & \textbf{\tool} & \textbf{Upper Bound} & \textbf{\tool} & \textbf{Upper Bound}\\
\midrule
1 & 108~(36.0\%) & 108~(36.0\%) & 142~(47.3\%) & 142~(47.3\%) \\
2 & 116~(38.7\%) & 125~(41.7\%) & 148~(49.3\%) & 151~(50.3\%) \\
4 & 126~(42.0\%) & 136~(45.3\%) & 161~(53.7\%) & 167~(55.7\%) \\
6 & 138~(46.0\%) & 143~(47.7\%) & 175~(58.3\%) & 182~(60.7\%) \\
\bottomrule
\end{tabular}
\caption{Effect of Candidate Patch Pool Size on KGCompass Repair Accuracy}
\label{tab:top6}
\end{table}

\find{{\bf [RQ-3] Findings:} (1) The hybrid KG+LLM locator delivers the highest coverage. With Claude-4, it reaches 91.0\% file-level and 66.7\% function-level, and with Claude-3.5, it reaches 84.3\% and 57.7\%. (2) KG-derived entity paths improve repair success beyond KG-driven location alone, from 131 to 142 resolved on Claude-4 and from 102 to 108 on Claude-3.5 at temperature $0.0$. (3) The layered ranking pipeline raises success to 175 (58.3\%) on Claude-4 and 138 (46.0\%) on Claude-3.5, both close to their upper bounds. (4) Leveraging the ground-truth upper bound, \tool with Claude-4 reaches 182 (60.7\%), higher than ExpeRepair’s 181 (60.3\%), indicating meaningful headroom for stronger ranking. {\bf Insights:} (1) Knowledge graph not only improves location, but also guides the LLM's patch generation through entity paths; the same idea could be applied to other LLM-based tasks. (2) Patch selection strongly affects final repair success; systematic yet lightweight ranking rules already approach the upper limit, so future work on learning-based or more robust rankers may close the remaining performance gap without manual design and heavy cost.}

\section{Threats to Validity}

\paragraph{Internal Validity}
Internal validity concerns potential experimental biases that could affect the fairness of the results. To prevent this, we enforced strict temporal constraints during our knowledge graph construction, incorporating only issues and pull requests that existed before each issue's creation time within SWE-bench Lite, thereby preventing data contamination from future artifacts. Similarly, for patch evaluation, we used only regression tests that passed successfully in the original codebase before the bug was reported, ensuring a fair assessment of repair capabilities.

\paragraph{External Validity}
External validity addresses how far our conclusions extend beyond the Python projects in SWE-Bench Lite. \tool mitigates this threat by keeping language-specific logic in thin adapter layers. Each language implements a lightweight \texttt{LanguageConfig} and a corresponding \texttt{Parser}, while graph construction, ranking, and repair modules remain unchanged. Adding support for other programming languages, therefore, requires only modest, self-contained code, leaving the core workflow intact. This modular design reduces dependence on any single language and facilitates adoption in other systems.

\paragraph{Construct Validity}
Construct validity concerns whether we accurately measure what we claim to evaluate. To address this, we employed multiple complementary metrics: repair success rate (``Resolved''), fault location accuracy at file and function levels (``File Acc.'' and ``Func Acc.''), and computational efficiency (``Avg Cost''). By comparing \tool's generated patches with ground truth patches, we provided validation beyond test-passing success, helping identify truly correct fixes rather than merely coincidental solutions that happen to pass tests.

\section{Related Work}
\subsection{Repository-level Software Repair}
Recent large language models (LLMs) have demonstrated remarkable capabilities in repository-level software repair tasks~\cite{xia2024agentless,zhang2024autocoderover,wang2024openhandsopenplatformai,morepair}. SWE-Bench Lite~\cite{jimenez2024swebench} provides a representative benchmark of 300 real-world GitHub issues, making it suitable for evaluating LLM-based software repair approaches. Repository-level repair approaches fall into two categories: \textbf{Agentic approaches} coordinate multiple specialized agents through observe-think-act loops~\cite{wang2024openhandsopenplatformai,zhang2024autocoderover,ruan2024specrover,yang2024sweagent,koduai,composio}. OpenHands~\cite{wang2024openhandsopenplatformai} employs agents interacting with sandbox environments, while AutoCodeRover~\cite{zhang2024autocoderover,ruan2024specrover} leverages hierarchical code search for repository navigation. Composio~\cite{composio} implements a state-machine multi-agent system using LangGraph~\cite{LangGraph2024}, and SWE-agent~\cite{yang2024sweagent} introduces specialized tools for LLM interactions. \textbf{Procedural approaches} employ expert-designed sequential workflows for repair~\cite{xia2024agentless,zhang2024autocoderover,antoniades2024swe,gautam2024supercoder2}. Agentless~\cite{xia2024agentless} employs a three-phase repair, including fault location, patch generation, and patch validation. Moatless~\cite{antoniades2024swe,orwall2024moatless} proposed that, rather than relying on an agent to reason its way to a solution, it is crucial to build good tools to insert the right context into the prompt and handle the response.

Pure LLM-based approaches treat repository artifacts and code as separate entities, rather than as interconnected components, which hinders their ability to bridge the gap between natural language issues and the structured codebase. In contrast, \tool explicitly models both the structural codebase and the semantic relationships across repository artifacts through knowledge graphs.

\subsection{Knowledge Graph for Repository-level Software}
Knowledge graphs have emerged as a powerful approach for modeling code repositories by capturing complex relationships between code entities and their dependencies~\cite{zhao2019knowledge,liu2024graphcoder,ouyang2024repograph,liu2024codexgraph}.
Recent work has demonstrated the effectiveness of knowledge graphs in understanding software repositories. RepoGraph~\cite{ouyang2024repograph} operates at a fine-grained code line level, demonstrating that modeling explicit definition-reference relationships can effectively guide repair decisions. GraphCoder~\cite{liu2024graphcoder} introduces control flow and dependency analysis into code graphs, showing that structural code relationships are critical for understanding code context and predicting the subsequent statements in completion tasks. CodeXGraph~\cite{liu2024codexgraph} represents code symbols and their relationships in graph databases, proving that querying graph structures enables more precise retrieval of relevant code snippets than traditional sequence-based approaches.

Existing graph-based approaches primarily focus on code structure, overlooking key \textbf{repository artifacts} (such as issues and pull requests), and fail to \textbf{leverage indirect entity relationships} for patch generation. \tool integrates repository artifacts, combining codebase entities, into the knowledge graph and introduces path-guided repair, strengthening the link between issues and relevant functions and enabling more precise repairs.
\section{Conclusion}
This paper introduces \tool, a repository-aware repair framework that constructs a knowledge graph linking issues, pull requests, files, and functions, then steers a large language model along ranked multi-hop paths to generate patches. On the SWE-Bench Lite benchmark, \tool repairs 58.3\% of bugs and achieves 83.6\% file-level and 56.0\% function-level fault location at an average cost of \$0.2 per instance, outperforming state-of-the-art single-LLM baselines, and boosting repair success by 30.2\% to 156.4\% across four backbone LLMs. The graph reduces the search space from thousands to 20 candidates, addressing the context-length constraint that limited repository-level repair. 89.7\% of successful fault locations require 2+ hops; therefore, the knowledge graph successfully bridges the semantic gap between natural-language bug reports and codebase by capturing indirect structural relations that text-only LLMs miss. Path-guided prompting attaches an explicit reasoning chain to each candidate, and ablations show that omitting it sharply lowers accuracy. Supplying the language model with this compact, context-rich candidate set further stabilizes repair quality and reduces inference cost, thereby mitigating the model-dependent variability observed in pure LLM workflows. Future work could enrich the knowledge graph with domain-specific knowledge, and could extend the path-guided paradigm to multi-hop question answering, fact verification, and other natural-language reasoning tasks.



\bibliographystyle{ACM-Reference-Format}
\bibliography{sample-base}

\end{document}